\documentclass[preprint,showpacs,aps,nofootinbib]{revtex4}
\usepackage{graphicx}
\usepackage{amsmath}
\usepackage{bm}

\begin{document}
\preprint{ND Atomic Theory 2003-1}
\title{Combined effect of coherent $Z$ exchange and
the hyperfine interaction in
atomic  PNC}
\author{W. R. Johnson}
 \affiliation{Institute for Nuclear Theory, University of Washington\\
  Seattle, WA 98195}
 \affiliation{Department of Physics, 225 Nieuwland Science Hall\\
  University of Notre Dame, Notre Dame, IN 46566}
 \email{johnson@nd.edu}
 \homepage{http://www.nd.edu/~johnson}
\author{M. S. Safronova}
 \email{msafrono@nd.edu}
 \affiliation{Electron and Optical Physics Division,
  National Institute of Standards and Technology\\
  Gaithersburg, MD, 20899}
\author{U. I. Safronova}
 \email{usafrono@nd.edu}
 \affiliation{Department of Physics, 225 Nieuwland Science Hall\\
   University of Notre Dame, Notre Dame, IN 46566}
\date{\today}
\begin{abstract}
The nuclear spin-dependent parity nonconserving (PNC) interaction
arising from a combination of the hyperfine interaction and the
coherent, spin-independent, PNC interaction from $Z$ exchange is
evaluated using many-body perturbation theory. For the
$6s_{1/2}-7s_{1/2}$ transition in $^{133}$Cs, we obtain a result
that is about 40\% smaller than that found previously by Bouchiat
and Piketty [Phys.\ Lett.\ B {\bfseries 269}, 195 (1991)].
Applying this result to $^{133}$Cs, leads to an increase in the
experimental value of nuclear anapole moment and exacerbates
differences between constraints on PNC meson coupling constants
obtained from the Cs anapole moment and those obtained from other
nuclear parity violating experiments. Nuclear spin-dependent PNC
dipole matrix elements, including contributions from the combined
weak-hyperfine interaction, are also given for the
$7s_{1/2}-8s_{1/2}$ transition in $^{211}$Fr and for transitions
between ground-state hyperfine levels in K, Rb, Cs, Ba$^{+}$, Au,
Tl, Fr, and Ra$^{+}$.

\end{abstract}
\maketitle

\section{Introduction\label{sec1}}

The precise measurements of the $6s[F\! =\! 4] - 7s[F\! =\! 3]$
and $6s[F\! =\! 3] - 7s[F\! =\! 4]$ parity nonconserving (PNC)
dipole matrix elements in $^{133}$Cs by \citet{wood:97} lead to a
value of the weak charge\footnote{The experimental value includes
a net correction of -1.1\% to the theoretical PNC amplitude
\cite{DFS:89,BSJ:92,DFG:02} from the Breit interaction
\cite{Der:00}, $\alpha Z$ vertex corrections \cite{KF:02,MST:02},
Coulomb-field vacuum polarization \cite{JS:01}, and nuclear skin
effects \cite{PW:99,JS:99}. } $Q^{\text{exp}}_W(^{133}\text{Cs})
= -72.73(46)$ that is in agreement with the Standard Model value
$Q^{\text{SM}}_W(^{133}\text{Cs}) = -73.09(3)$ \cite{19}. These
measurements also lead to an experimental value of the much
smaller contribution from the nuclear spin-dependent PNC
interaction that is accurate to about 15\%. This spin-dependent
contribution has three distinct sources: the nuclear anapole
moment \cite{ana:58,flam:84},  the
 $Z$ exchange interaction from nucleon axial-vector ($A_n V_e$)
currents, and the combined action of the hyperfine interaction and
the spin-independent $Z$ exchange interaction from nucleon vector ($V_n A_e$)
currents \cite{FK:85,BP:91a}. Of these three, the anapole contribution
dominates.
The contributions from the
anapole and nuclear axial-vector current are
\begin{equation}
H^{(i)} = \frac{G}{\sqrt{2}}\ \kappa_{i}\ \bm {\alpha \cdot I}\ \rho(r),
\label{eq1}
\end{equation}
where $G$ is the universal weak coupling constant, $\bm{I}$ is the
nuclear spin, and $\rho(r)$ is a normalized nuclear density
function. The subscript $i$ of the dimensionless constants
$\kappa_i$ takes the values $i=a$ for the anapole contribution and
$i=2$ for the axial-vector contribution. In Refs.~\cite{FK:85,BP:91a},
the hyperfine--vector current contribution was also reduced to
the form given in Eq.~(\ref{eq1}) with a corresponding dimensionless
constant $\kappa_\text{hf}$.

To extract the anapole contribution $\kappa_a$ from
experiment, it is necessary to know the corresponding spin-dependent
PNC amplitude calculated with $\kappa_a=1$, as well as the
two contributions from the axial-vector and weak-hyperfine interference
terms quantified by $\kappa_2$ and $\kappa_\text{hf}$. The spin-dependent
PNC amplitude was calculated in various approximations in
Refs.~\cite{ana:88,ana:88a,BSJ:92}. Nuclear shell-model values of
$\kappa_2$ for $^{133}$Cs and $^{203}$Tl were obtained in recent
 calculations by
\citet{HLR:01}.  An analytical approximation for $\kappa_\text{hf}$
was derived by \citet{FK:85} and values of $\kappa_\text{hf}$ were
later determined for various cases of experimental interest by
\citet{BP:91a}.

Recently, \citet{HAX:01} used the values of
$\kappa_2$ and $\kappa_\text{hf}$\footnote{In Ref.~\cite{HAX:01},
$\kappa_2$ and $\kappa_\text{hf}$ are designated by $\kappa_{Z_0}$
and $\kappa_{Q_W}$, respectively.} determined as described
above to extract values of
$\kappa_a$ from PNC measurements in $^{133}$Cs \cite{wood:97}. The resulting
anapole moments were, in turn, used to place constraints on PNC
meson coupling constants \cite{DDH:80}. The constraints obtained
from the Cs experiment were found to be inconsistent with
constraints from other nuclear PNC measurements, which favor a
smaller value of the $^{133}$Cs anapole moment.

Motivated by this disagreement, we are led to re-examine the
combined hyperfine-weak interaction. We find that the
contribution of this term to the PNC dipole matrix element at the
Dirac-Hartree-Fock (DHF) level can be approximated
by a spin-dependent interaction of
the type given in Eq.~(\ref{eq1}); however, such an approximation
is not justified in correlated calculations, since contributions
from Eq.~(\ref{eq1}) are very sensitive to correlations, whereas
contributions from the combined hyperfine-weak interaction are
relatively insensitive to correlation corrections. We do find,
nevertheless, that even in correlated calculations there is a
rough proportionality between contributions from the combined
interaction and those from the interaction given in
Eq.~(\ref{eq1}) that is independent of hyperfine state, and we use
this fact to define  ``effective'' values of the coupling strength
$\kappa_\text{hf}$ for cases of potential experimental interest. For
the $6s-7s$ transition in Cs, our effective value of $\kappa_\text{hf}$
is about 40\% smaller than the value from  \cite{BP:91a}. Interestingly,
for this case our final correlated value of $\kappa_\text{hf}$
is quite close to the value predicted by the formula derived in \cite{FK:85}.
Other things being unchanged, this decrease in the size of $\kappa_\text{hf}$
leads to an
increase in the size of $\kappa_a$ and, correspondingly, in the Cs
anapole moment; consequently, increasing the inconsistencies
between various experimental constraints on PNC meson coupling constants
described in \cite{HAX:01}.

\section{Method}

We write the spin-dependent PNC correction to the reduced electric-dipole
matrix element $\langle w F_F\|z\|v F_I \rangle$ as the sum of three
terms:
\begin{equation}
\langle w F_F\|z\|v F_I \rangle^{\text{sd}}_{\text{PNC}} =
\kappa_a \: \langle w F_F\|z\|v F_I \rangle^{(a)}+ \kappa_{2}
\:\langle w F_F\|z\|v F_I \rangle^{(2)}+ \langle w F_F\|z\|v F_I
\rangle^{(\text{hf})},
\end{equation}
where indices $(a)$, $(2)$, and $(\text{hf})$ correspond to
the anapole, axial-vector, and weak-hyperfine
interference, respectively.
Since the anapole and axial-vector contributions both
take the form given in Eq.~(\ref{eq1}), we can introduce
$\langle w  F_F\|z\|v F_I \rangle^{(2,a)}\equiv\langle w F_F\|z\|v F_I
\rangle^{(a)}=\langle w F_F\|z\|v F_I \rangle^{(2)}$.
We then define  $\kappa_\text{hf}$ as the ratio
\begin{equation}
 \kappa_\text{hf}=\frac{\langle w F_F\|z\|v F_I \rangle^{(\text{hf})}}{\langle w F_F\|z\|v F_I
 \rangle^{(2,a)}}.
 \end{equation}
We expect, and indeed find, that $\kappa_\text{hf}$ depends on the initial and final
hyperfine levels. For cases considered here, however, the dependence of $\kappa_\text{hf}$
on the hyperfine levels $F_I$ and $F_F$ is weak and we may treat $\kappa_\text{hf}$
as constant to some level of accuracy.
We write the expression for the total spin-dependent PNC
contribution to the electric-dipole
matrix element $\langle w F_F\|z\|v F_I \rangle$  as
 \begin{equation}
 \langle w F_F\|z\|v F_I \rangle^{\text{sd}}_{\text{PNC}} =
(\kappa_{a}+\kappa_{2}+\kappa_\text{hf}) \langle w F_F\|z\|v F_I
\rangle^{(2,a)}.  \label{e4}
\end{equation}
 and define
\begin{equation}
\kappa=\kappa_{a}+\kappa_{2}+\kappa_{\text{hf}}. \label{kappa}
\end{equation}
In this work, we calculate both $\langle w F_F \|z\|v F_I
\rangle^{(\text{hf})}$ and $\langle w F_F \|z\|v F_I \rangle^{(2,a)}$
and, therefore, determine the state-dependent values of $\kappa_\text{hf}$.

The hyperfine interaction Hamiltonian is written
\begin{equation}
H^{\text{(\text{hf})}} = -ec\ \bm{\alpha \cdot A},\quad \bm{A}(\bm{r}) = \frac{\mu_0}{4\pi}\
\int\!\! d^3 x \ \frac { {\bm M}(x) {\bm \times} ({\bm r}-{\bm x})}{|{\bm r}-{\bm x}|^3} ,
\label{eqm}
\end{equation}
where ${\bm M}(x)$ is the magnetization density, which is related
to the nuclear moment ${\bm \mu}_I$ by
\[
{\bm \mu}_I = \int\!\! d^3 x \ {\bm M}(x) =  g_I {\bm I} \mu_N.
\]
Here $\mu_N = |e|\hbar/2M_p$ is the nuclear magneton.

The dominant, spin-independent, part of the weak interaction is
\begin{equation}
H^{(1)} = \frac{G}{2\sqrt{2}}\ Q_W\ \gamma_5\ \rho(r) , \label{yy}
\end{equation}
where $Q_W$ is the conserved weak charge of the nucleus, given at tree level in terms of the
neutron number $N$, the proton number $Z$, and the Weinberg angle
$\theta_W$ by $Q_W= Z(1-4\sin^2{\theta_W}) -N$, and $\rho(r)$ is a nucleon distribution
function.
In our numerical calculations of the interference term,
we use radiatively corrected values of $Q_W$ inferred from \cite{19}.
The nucleon
distribution $\rho(r)$ is assumed to have the form
\begin{equation}
 \rho(r) = \frac{\rho_0}{1+\exp{[(r-C)/a]}} \label{rho}
\end{equation}
where $a = 0.523$ fm (corresponding to 90\%--10\% fall-off thickness $t=2.3$~fm)
and where $C$ is inferred from
the nuclear charge radii listed in \cite{JS:85}.
In the exceptional case of
$^{211}$Fr, we choose $C = 6.733$~fm, corresponding to the value
$R_{\text{rms}} = 5.566$~fm given in Ref.~\cite{Fr:99}.
We assume that the radial dependence of the
magnetization distribution is identical to that of the nucleon
distribution.

As shown in Appendix \ref{A},
the dipole matrix element corresponding to the weak-hyperfine interference is given by the
third-order perturbation theory expression
\begin{multline}
\langle w I F_F M_F | z | v I F_I M_I \rangle^\text{(hf)} = \\
\quad \sum_{ \substack{m\neq w\\ n\neq w} }
\frac{\langle w | H^{(1)} | n\rangle
\langle n| H^\text{(hf)} | m \rangle \langle m | z | v\rangle}{(E_w-E_m)(E_w-E_n)}
+
\sum_{ \substack{m\neq w\\ n\neq w} }
\frac{\langle w | H^\text{(hf)} | n\rangle
\langle n| H^{(1)} | m \rangle \langle m | z | v\rangle}{(E_w-E_m)(E_w-E_n)}\\
+ \sum_{\substack{m\neq w\\ n\neq v}}
\frac{\langle w | H^{(1)}  | m \rangle \langle m | z | n \rangle
\langle n | H^\text{(hf)} | v \rangle}
{(E_w-E_m)(E_v-E_n)}
+ \sum_{\substack{m\neq w\\ n\neq v}}
\frac{\langle w |  H^\text{(hf)} | m \rangle \langle m | z | n \rangle
\langle n |  H^{(1)} | v \rangle}
{(E_w-E_m)(E_v-E_n)}
\\
+
\sum_{ \substack{m\neq v\\ n\neq v} }
\frac{\langle w | z | n\rangle
\langle n| H^{(1)} | m \rangle \langle m | H^\text{(hf)} | v\rangle}{(E_v-E_m)(E_v-E_n)}
+ \sum_{ \substack{m\neq v\\ n\neq v} }
\frac{\langle w | z | n\rangle
\langle n| H^\text{(hf)}| m \rangle \langle m | H^{(1)} | v\rangle}{(E_v-E_m)(E_v-E_n)}\\
- \langle w |H^\text{(hf)} | w\rangle \sum_{m\neq w}
\frac{\langle w | H^{(1)} | m \rangle \langle m | z | v \rangle}{(E_w-E_m)^2}
- \sum_{n\neq v}
\frac{\langle w | z | n \rangle\langle n | H^{(1)} | v \rangle }{(E_v-E_n)^2}\
\langle v |H^\text{(hf)} | v\rangle,\label{eq2}
\end{multline}
where we use designations $|w\rangle$ and $|v\rangle$ on the right-hand side for
coupled hyperfine states $|w I F_F M_F\rangle$ and $|v I F_I M_I\rangle$,
respectively, and where we designate the energy of state $i$ by $E_i$.
Note that the other matrix element $\langle wI F_F M_F | z | v I F_I M_I  \rangle^{(2,a)}$
in Eq.~(\ref{e4}) is obtained from a considerably simpler
second-order perturbation theory calculation.

In Ref.~\cite{BP:91a}, terms on the second and fourth lines of Eq.~(\ref{eq2}) were
ignored and
partial sums on the first and third lines, such
as
\[
\sum_{n\neq v} \frac{ H^{(1)} | n \rangle\ \langle n | H^{\text{(\text{hf})}} }
{E_n-E_v},
\]
were carried out using
free-particle Green's functions and reduced to an effective interaction
\[
H^{(\text{eff})} = \frac{G}{\sqrt{2}}\ \kappa_\text{hf}\ \bm {\alpha \cdot I}\ \rho(r)
\]
of the form given in Eq.~(\ref{eq1}).
A similar reduction was made in Ref.~\cite{FK:85}.
Here, we evaluate all of the terms in the Eq.~(\ref{eq2}) numerically.
One important advantage of this direct numerical evaluation
is that correlation corrections to $\langle w I F_F\| z \| v I F_I\rangle^{(\text{hf})}$
can be determined
using standard many-body methods. Indeed, we find that Eq.~(\ref{eq2}) is
insensitive to correlations at the random-phase approximation (RPA) level
for most of the cases considered here, whereas calculations based on the
contracted approximation above
are very sensitive to correlation corrections!
A reduction of Eqs.~(\ref{e4}) and (\ref{eq2}) to reduced matrix elements
suitable for numerical evaluation is given in Appendix~\ref{B}.

\section{Numerical Results}

We evaluate the reduced dipole matrix element
$\langle w F_F\| z \| v F_I\rangle^{(\text{hf})}$ given in (\ref{eq2p3}) and
the reduced matrix element $\langle w F_F\| z \| v F_I\rangle^{(2,a)}$
given in (\ref{eq3p3}),
and we find that their ratio is approximately independent of the angular momentum
quantum numbers $F_I$ and $F_F$ for transitions between hyperfine levels.

Let us consider the $6s-7s$ transition in Cs.  We first evaluate the reduced
matrix elements in (\ref{eq3p3}) and (\ref{eq2p3}) at the DHF level of approximation.
We solve the DHF equations in a finite B-spline basis using the methods described in
\cite{JBS:88} and use the resulting basis functions to evaluate matrix elements and carry out
sums over intermediate states.
For the case Cs, our basis set consists of 100 splines of order 15 for each angular momentum
state. The basis orbitals are constrained to a cavity of radius 45 a.u.;
the cavity radius is modified in other atoms to accommodate the initial and final
valence orbitals. As a check, we carried out the Cs calculations
using a cavity of radius 75 a.u.\ to verify that the results are stable against
changes in the cavity radius.
Results of our DHF calculations for the transitions between the possible hyperfine levels
are presented in the upper four rows of Table~\ref{tab1}.  We find that the ratio
$\kappa_\text{hf}$ of the $\langle w F_F\| z \| v F_I\rangle^{(\text{hf})}$ to
$\langle w F_F\| z \| v F_I\rangle^{(2,a)}$ matrix element changes from level to level
by only 2\% in the DHF approximation.

\begin{table}
\caption{We list values of $\kappa_\text{hf}$ for
transitions between hyperfine levels $6s[F_I]-7s[F_F]$ in Cs determined in DHF and RPA
approximations.
The atomic number is $A=133$, the nuclear spin is $I=7/2$, the
nuclear magnetic moment $\mu_I= 2.5826$,
the weak charge (including radiative corrections) is $Q_W= -73.09(3)$,
and  the 50\%
fall off radius is $C= 5.675$ fm for both the nuclear $\rho(r)$ and magnetization $M(r)$
distributions;
the 10\%-90\% fall off distance is 2.3 fm.  The PNC reduced dipole matrix
elements $\langle w F_F\| z \| v F_I\rangle^{(2,a)}$ are given
together with the weak-hyperfine interference correction
to dipole matrix elements $\langle w F_F|| z || v F_I\rangle^{(\text{hf})}$;
their ratio is $\kappa_\text{hf}$.
Numbers in square brackets represent powers of 10.\label{tab1}}
\begin{ruledtabular}
\begin{tabular}{lcccc}
\multicolumn{1}{c}{Type} & \multicolumn{1}{c}{$F_F-F_I$} &
\multicolumn{1}{c}{$\langle 7s F_F|| z || 6s F_I \rangle^{(2,a)}$} &
\multicolumn{1}{c}{$\langle 7s F_F|| z || 6s F_I\rangle^\text{(hf)}$} &
\multicolumn{1}{c}{$\kappa_\text{hf}$}\\ \hline
    DHF    & 3 - 3&  1.908[-12]& 1.193[-14]  &  6.251[-03]    \\
    DHF    & 3 - 4&  5.481[-12]& 3.480[-14]  &  6.349[-03]    \\
    DHF    & 4 - 3&  4.746[-12]& 3.020[-14]  &  6.364[-03]    \\
    DHF    & 4 - 4&  2.173[-12]& 1.358[-14]  &  6.251[-03]    \\[0.8ex]
    RPA    & 3 - 3&  2.249[-12]& 1.141[-14]  &  5.076[-03]    \\
    RPA    & 3 - 4&  7.299[-12]& 3.579[-14]  &  4.903[-03]    \\
    RPA    & 4 - 3&  6.432[-12]& 3.139[-14]  &  4.880[-03]    \\
    RPA    & 4 - 4&  2.560[-12]& 1.300[-14]  &  5.076[-03]    \\[0.2ex]
\end{tabular}
\end{ruledtabular}
\end{table}

\begin{table}
\caption{We list  values of
$\kappa_\text{hf}$ for transitions between hyperfine levels
$8s[F_F]-7s[F_I]$ in Fr determined in RPA type
calculations. Here, the atomic number $A=211$, the nuclear spin is
$I=9/2$, the nuclear magnetic moment $\mu_I= 4.00$, the weak
charge (including radiative corrections) is $Q_W= -116.23$, and
the 50\% fall off radius is $C= 6.7325$ fm for both the nuclear
$\rho(r)$ and magnetization $M(r)$ distributions; the 10\%-90\%
fall off distance is 2.3 fm. The PNC reduced dipole matrix
elements $\langle w F_F\| z \| v F_I\rangle^{(2,a)}$  are given
together with the weak-hyperfine interference correction
 to dipole matrix elements
$\langle w F_F|| z || v F_I\rangle^{(\text{hf})}$;
their ratio is $\kappa_\text{hf}$.
 Numbers in square brackets
represent powers of 10.\label{tab1fr}}
\begin{ruledtabular}
\begin{tabular}{cccc}
 \multicolumn{1}{c}{$F_F-F_I$} &
\multicolumn{1}{c}{$\langle 8s F_F|| z || 7s F_I \rangle^{(2,a)}$} &
\multicolumn{1}{c}{$\langle 8s F_F|| z || 7s F_I\rangle^{(\text{hf})}$} &
\multicolumn{1}{c}{$\kappa_\text{hf}$}\\ \hline
4 - 4& 3.092[-11]&3.472[-13] & 1.123[-02]\\
4 - 5& 1.016[-10]&1.069[-12] & 1.053[-02]\\
5 - 4& 9.224[-11]&9.645[-13] & 1.046[-02]\\
5 - 5& 3.426[-11]&3.846[-13] & 1.123[-02]\\
\end{tabular}
\end{ruledtabular}
\end{table}

The DHF treatment of PNC in cesium is known to be a rather poor approximation, giving
a value for the dominant part of the PNC dipole matrix element that is 20\% smaller
than the final correlated value. To obtain a reliable value for the PNC matrix element,
one must go beyond the DHF approximation and treat correlation corrections. The dominant
correlation corrections, those associated with core shielding, are obtained in the
random-phase approximation. Including RPA corrections to both weak-interaction
and dipole
matrix elements gives a value for the dominant PNC dipole matrix element in Cs that
is within 2\% of the final
correlated value.

The RPA matrix elements are calculated as
described in Ref.~\cite{ADNDT:96}, with the value of  $\omega$ in the RPA
equations set to zero.
In the last four rows of Table~\ref{tab1},
we give values of $\langle 7s F_F|| z || 6s F_I \rangle^{(2,a)}$
that include RPA corrections to both dipole and weak-interaction
operators, and values of $\langle 7s F_F|| z || 6s F_I\rangle^{(\text{hf})}$
that include RPA corrections to the dipole,
weak-interaction, and hyperfine operators. While the RPA values of
$\langle 7s F_F|| z || 6s F_I \rangle^{(2,a)}$ are 15-25\% larger
than the DHF values, the RPA and DHF values of
$\langle 7s F_F|| z || 6s F_I\rangle^{\text{(hf)}}$
differ by only 3-5\%. Thus, by contrast to PNC
dipole matrix elements induced by the dominant spin-independent
interaction and by the spin-dependent interactions given in
Eq.~(\ref{eq1}), which are very sensitive to correlation
corrections, the third-order matrix elements for the combined
interaction are relatively insensitive to correlations
for the $6s-7s$ transition in Cs.  It should be
emphasized that the contraction of operators introduced in
\cite{BP:91a} is a useful approximation at the
independent-particle DHF level of approximation; however, when
correlation corrections are included, although an approximate
proportionality still obtains, the proportionality {\it constant}
depends on correlations; this is a reflection of the fact that
there is no effective Hamiltonian of form (\ref{eq1}) for the combined
interaction.

We include negative-energy contributions \cite{SDB:99} when evaluating sums
over intermediate states in Eqs.~(\ref{eq2p3}) and (\ref{eq3p3})
and when calculating RPA matrix elements.  We find almost
no negative-energy correction to $\langle 7s F_F|| z || 6s F_I \rangle^{(2,a)}$.
However, the negative-energy corrections to
$\langle 7s F_F|| z || 6s F_I\rangle^\text{(hf)}$
were found to be large, 22-23\% at both the DHF and RPA levels of approximation.
Since negative-energy contributions are important for accurate calculation of
$\langle 7s F_F|| z || 6s F_I\rangle^\text{(hf)}$, they are, therefore,
important in the evaluation of $\kappa_\text{hf}$.
Omission of negative-energy contributions leads to values of
$\kappa_{\text{hf}}=0.0049$ in the DHF approximation and
$\kappa_{\text{hf}}=0.0038$ in the RPA approximation which are about 20\%
smaller than our final values listed in Table~\ref{tab1}. We note that
our final correlated value fortuitously coincides with the
DHF value without
negative energy contributions; the correlation correction
decreases the value of $\kappa_{\text{hf}}$ and the negative energy
contribution increases $\kappa_{\text{hf}}$ by
approximately the same amount.
We stress again that these two effects contribute, in fact, to different
quantities, negative energies contribute only
to $\langle 7s F_F|| z || 6s F_I \rangle^{\text{(hf)}}$ and correlation
primarily to $\langle 7s F_F|| z || 6s F_I \rangle^{(2,a)}$.

We also found that sums in the interference matrix element given in Eq.~(\ref{eq2p3})
must include the entire set of basis orbitals, in contrast to
the sums in Eq.~(\ref{eq3p3}), where omitting  high-energy orbitals
from the basis has very little effect. In other words, the completeness
of the basis is very important for calculation of the sums in (\ref{eq2p3}).

Results of our RPA calculations for the $7s-8s$
transitions in Fr are presented in Table~\ref{tab1fr}.
The state dependence of $\kappa_\text{hf}$
increases to 6-7\% in Fr in comparison to Cs, where differences
in $\kappa_\text{hf}$ for different transitions were 3-4\%
in the RPA approximation.
As in the case of Cs, the largest differences occur between transitions
with $F_I=F_F$ and those with $F_F\neq F_I$; there is only 0.7\%
difference in $\kappa_\text{hf}$ between the 4-5 and 5-4 transitions.

For the 4--3 and 3--4 hyperfine transitions in Cs measured by \citet{wood:97},  an
effective value
$\kappa_{\text{hf}}= 0.0049$ can be extracted from  the RPA values listed in
Table~\ref{tab1}. This value is about 40\%
smaller than the value $\kappa_{\text{hf}}= 0.0078$  from \cite{BP:91a}
but agrees exactly with the value obtained earlier by \citet{FK:85}.
We use our value of $\kappa_{\text{hf}}$ to extract a
value of $\kappa_a$ from the  Cs PNC experiment of \citet{wood:97},
  \begin{equation}
\Delta\left[\frac{\textrm{Im}(E_{\textrm{PNC}})}{\beta}\right]_{34-43}=
- 0.077\pm 0.011 \: \text{mV/cm} , \label{expt}
\end{equation}
where $\beta$ is the vector polarizability of the transition, which
has been measured in Ref.~\cite{BW:99} with high accuracy $\beta=27.02(8)~a^3_0$.
The subscripts $34$ and $43$ in (\ref{expt}) correspond to $F_F\, F_I$.

 The spin-independent PNC amplitude $E^{(1)}_\text{PNC}$ in
alkali-metal atoms ($j_F=j_I=1/2$) is customarily defined as
\begin{equation}
E^{(1)}_{\text{PNC}}=\langle j_F\ 1/2|z|j_I\ 1/2 \rangle  ,
\end{equation}
 leading to the following relation between
spin-dependent PNC amplitude and the corresponding spin-dependent
reduced matrix elements:
\begin{equation}
E^{\text{(sd)}}_\text{PNC} = \frac{\kappa}{\cal A}\:
 \langle w F_F \|z\|v F_I \rangle^{\text{(2,a)}},
\end{equation}
where the $\kappa$ is defined by Eq.~(\ref{kappa}) and ${\cal A}$ is
an angular coefficient
\begin{equation}
{\cal A} =(-1)^{j_F+F_I+I+1}\
\sqrt{6\,\left[F_I \right]\left[F_F \right]}\left\{
\begin{array}{ccc}
F_F & F_I & 1 \\
 j_I & j_F & I
\end{array}
\right\},
\end{equation}
where $[F]=2F+1$.
For the two transitions considered here,
 ${\cal A}_{43}=-{\cal A}_{34}$, so we may write
 \begin{equation}
 \Delta\left[\frac{\textrm{Im}(E_{\textrm{PNC}})}{\beta}\right]_{34-43}=
  - \frac{\kappa}{{\cal A}_{43}\, \beta}
  \left[\langle 7s F_F \|z\|6s F_I \rangle^{(2,a)}_{34}+\langle 7s F_F \|z\|
  6s F_I  \rangle^{(2,a)}_{43}\right] \frac{e}{4\pi \epsilon_0 a^2_0}
  \label{kappa1}
\end{equation}

\begin{table}
\caption{Comparison of contributions to spin-dependent PNC in
$^{133}$Cs obtained by different groups.
All results are presented in terms of the coefficients
 $\kappa_a$, $\kappa_2$, $\kappa_{\text{hf}}$, and their sum $\kappa$,
used in the present paper.   \label{comp}}
\begin{ruledtabular}
\begin{tabular}{lcccc}
   \multicolumn{1}{c}{Group} &
   \multicolumn{1}{c}{$\kappa$} &
    \multicolumn{1}{c}{$\kappa_2$} &
\multicolumn{1}{c}{$\kappa_{\text{hf}}$}&
\multicolumn{1}{c}{$\kappa_a$}\\
 \hline
  Present                            &  0.117(16)&  0.0140\footnotemark[1]&  0.0049  &0.098(16)\\
  Haxton {\it et al.}
   \protect\cite{HAX:01,HLR:01,HLR:02}&        0.112(16)\footnotemark[2]&  0.0140& 0.0078\footnotemark[3] &  0.090(16)\\
 Flambaum and Murray
 \protect\cite{FLM:97}               &  0.112(16)\footnotemark[4]&0.0111\footnotemark[5]&
   0.0071\footnotemark[6]
&  0.092(16)\footnotemark[7]\\ Bouchiat and
Piketty \protect\cite{BP:91a,BP:91b}  &  &  0.0084&   0.0078&
\end{tabular}
\end{ruledtabular}
   \footnotetext[1]{Refs.~\protect\cite{HAX:01,HLR:01,HLR:02}.}
    \footnotetext[2]{Ref.~\protect\cite{FLM:97}.}
    \footnotetext[3]{Ref.~\protect\cite{BP:91a}.}
 \footnotetext[4]{The spin-dependent matrix
elements from \protect\cite{ana:88,ana:88a} are used.}
    \footnotetext[5]{Shell-model value with $\text{sin}^2\theta_W = 0.23$.}
    \footnotetext[6]{This value was obtained by scaling the analytical result
    from \protect\cite{FK:85} $\kappa_{\text{hf}}=0.0049$ by a factor 1.5.}
   \footnotetext[7]{Contains a
1.6\% correction for finite nuclear size; the raw value is
0.094(16).}
    \end{table}

Combining the experimental results for
$\Delta\left[\textrm{Im}(E_{\textrm{PNC}})/{\beta}\right]_{34-43}$
and $\beta$ with our values for the spin-dependent matrix elements
from Table~\ref{tab1}, we obtain $\kappa=0.117(16)$. The
uncertainty comes from the uncertainty in the experimental value
on the left side of Eq.~(\ref{kappa1}); the uncertainty in
 $\beta$ is negligibly small.  In Ref.~\cite{FLM:97},
 $\kappa=0.112(16)$ was obtained by combining the same
experimental data \cite{wood:97,BW:99} with spin-dependent
matrix elements from Refs.~\cite{ana:88,ana:88a}. This value
is also used in
Refs.~\cite{HAX:01,HLR:01,HLR:02}. Differences with our value of
$\kappa$ come only from differences in
$\langle 7s F_F \|z\|6s F_I \rangle^{(2,a)}$.

Combining the effective value for $\kappa_\text{hf}$ with the value
$\kappa_2=0.0140$ from \cite{HLR:01} and the  value $\kappa =
0.117(16)$ obtained above leads to $\kappa_a = 0.098(16)$, which
is 8\% larger than the value $\kappa_a = 0.090(16)$ obtained by
\citet{HAX:01} and 6\% larger than the value $\kappa_a =
0.092(16)$ from \citet{FLM:97}. To clarify the sources of these
differences, we compare our results with those from
Refs.~\cite{FLM:97,HAX:01,HLR:01,HLR:02,BP:91a,BP:91b} in Table~\ref{comp}. We
scaled the constants given in \cite{FLM:97,BP:91a,BP:91b}
to represent them in terms of the
coefficients  $\kappa$, $\kappa_a$, $\kappa_2$, $\kappa_\text{hf}$ used
here.

 The revised value of
$\kappa_\text{hf}$ and $\langle 7s F_F \|z\|6s F_I \rangle^{(2,a)}$
obtained in this work increases the value of the $^{133}$Cs
anapole moment, and thereby slightly increases the differences
between various experimental constraints on PNC meson coupling constants
discussed in \cite{HAX:01}. Since correlation
corrections to  $\langle 7s F_F \|z\|6s F_I \rangle^{(2,a)}$
are large, 25\% at RPA level, further
accurate calculations are clearly desirable.

Measuring the PNC electric-dipole transition between ground-state hyperfine levels
is a potentially fruitful method for obtaining experimental anapole moments for
atoms other than Cs. Schemes have been proposed to carry out such measurements
and calculations have been carried out for various atoms in Refs.~\cite{NK:75,LS:77,GE:88,DB:98,pk:01}.
The contribution of the spin-independent interaction from $Z$ exchange, which
dominates the PNC dipole matrix element between different atomic levels,
vanishes for the microwave transitions between hyperfine states of the
same level.

As an aid to the analysis of these microwave experiments,
we give reduced matrix elements  $\langle  F_F|| z ||  F_I\rangle^{(2,a)}$
induced by the spin-dependent interaction
of Eq.~(\ref{eq1}) together with
values of the third-order dipole matrix element
$\langle F_F || z || F_I \rangle^\text{(hf)}$
for atoms of potential experimental interest
in Table~\ref{tab2}. The corresponding calculations were carried out at the RPA
level of approximation.  The ratio of  matrix elements again gives
$\kappa_\text{hf}$.
The ground state configurations of the atoms listed in the
table are $ns_{1/2}$ or $np_{1/2}$, and
the hyperfine levels have angular momentum $F=I\pm1/2$.
For some of the atoms considered in Table~\ref{tab2}, RPA correlation
corrections to the weak-hyperfine interference matrix element are
no longer small; they contribute 20\% and 34\% to
$\langle F_F || z || F_I \rangle^\text{(hf)}$
for Au and Ra$^+$, respectively.

\begin{table}
\caption{We list values of  $\kappa_\text{hf}$ for microwave
transitions between ground-state hyperfine levels $F_F-F_I$
 in atoms of potential experimental
interest. In this table, $A$ is the atomic number, $I$ is the nuclear spin, $\mu_I$ is the
nuclear moment, $Q_W$ is the weak charge (including radiative corrections), $C$ is the 50\%
fall off radius of both the nuclear $\rho(r)$ and magnetization $M(r)$ distributions
(the 10\%-90\% fall off distance is taken as 2.3 fm). The ground state configurations of the
atoms considered here are $ns_{1/2}$ or $np_{1/2}$ and the hyperfine levels have angular
momentum $F=I\pm1/2$. The PNC reduced dipole matrix elements induced by the spin-dependent
Hamiltonian of Eq.~(\ref{eq1}), $\langle F_F|| z || F_I\rangle^{(2,a)}$,
are given together with the third-order dipole matrix elements
$\langle F_F|| z || F_I\rangle^{(\text{hf})}$;
their ratio gives $\kappa_\text{hf}$. These calculations are carried
out at the RPA level of approximation. Numbers in square brackets represent powers of 10.
\label{tab2}}
\begin{ruledtabular}
\begin{tabular}{lcclcccclll}
\multicolumn{1}{c}{Atom} &
\multicolumn{1}{c}{A}    &
\multicolumn{1}{c}{I}    & \multicolumn{1}{c}{$\mu_I$} &
\multicolumn{1}{c}{$Q_W$}& \multicolumn{1}{c}{$C$(fm)} &
\multicolumn{1}{c}{$nl$} &
\multicolumn{1}{c}{$F_I-F_F$} &
\multicolumn{1}{c}{$\langle F_F|| z || F_I\rangle^{(2,a)}$} &
\multicolumn{1}{c}{$\langle F_F|| z || F_I\rangle^{(\text{hf})}$} &
\multicolumn{1}{c}{$\kappa_\text{hf}$}\\
\hline
    K       & 39& 3/2& 0.39149&  -18.39& 3.611& $4s$& 1 - 2& -2.222[-13]& -1.113[-16]& 5.01[-04]\\
    K       & 41& 3/2& 0.21448&  -20.36& 3.611& $4s$& 1 - 2& -2.222[-13]& -6.753[-17]& 3.04[-04]  \\
    Rb      & 85& 5/2& 1.3534 &  -44.75& 4.871& $5s$& 2 - 3& -2.550[-12]& -5.432[-15]& 2.13[-03]\\
    Rb      & 87& 3/2& 2.7515 &  -46.73& 4.871& $5s$& 1 - 2& -1.363[-12]& -1.027[-14]& 7.54[-03]\\
    Cs      &133& 7/2& 2.5826 &  -73.09& 5.675& $6s$& 3 - 4& -1.724[-11]& -7.791[-14]& 4.52[-03]\\
    Ba$^{+}$&135& 3/2& 0.83863&  -74.01& 5.721& $6s$& 1 - 2& -6.169[-12]& -2.217[-14]& 3.59[-03]\\
    Ba$^{+}$&137& 3/2& 0.93735&  -75.98& 5.721& $6s$& 1 - 2& -6.169[-12]& -2.544[-14]& 4.12[-03] \\
    Au      &197& 3/2& 0.14816& -110.88& 6.554& $6s$ &1 - 2& -1.601[-11]& -1.912[-14]& 1.19[-03]\\
    Tl  &203& 1/2& 1.6222 & -114.69& 6.618& $6p_{1/2}$&0 - 1&-3.000[-11]& -3.437[-13]& 1.15[-02] \\
    Tl & 205& 1/2& 1.6382 & -116.66& 6.618& $6p_{1/2}$&0 - 1&-3.000[-11]& -3.531[-13]& 1.18[-02] \\
    Fr      &211& 9/2& 4.00   & -116.23& 6.733& $7s$& 4 - 5& -2.379[-10]& -2.223[-12]& 9.34[-03]\\
    Fr      &223& 3/2& 1.17   & -128.08& 6.834& $7s$ & 1 - 2&-5.820[-11]& -5.187[-13]& 8.91[-03]\\
    Ra$^{+}$&223& 3/2& 0.2705 & -127.02& 6.866& $7s$ & 1 - 2&-5.987[-11]& -1.258[-13]& 2.10[-03] \\
 \end{tabular}
\end{ruledtabular}
\end{table}

\section{Summary}

We have considered the PNC dipole matrix elements induced by the combined
hyperfine--weak interaction and found that they are, at the few percent level of accuracy,
proportional to the PNC dipole
matrix elements induced by the spin-dependent interaction of Eq.~(\ref{eq1}),
independent of $F_I$ and $F_F$, for transitions $F_I-F_F$ between hyperfine levels.
The proportionality is not the result of an
operator identity, but of similar angular momentum structures
for the respective matrix elements. By carrying out calculations at
the RPA level of approximation, which are expected to be accurate to a few percent,
we are able to extract effective coupling constants $\kappa_\text{hf}$
from the calculations.
Although the dominant matrix element $\langle wI F_F  \| z \| v I F_I  \rangle^{(2,a)}$
is sensitive to correlation corrections,
increasing by 10--30\% in Cs and Fr when correlation corrections are included,
the matrix element
$\langle wI F_F \| z \| v I F_I  \rangle^\text{(hf)}$
is correlation insensitive,
changing by less than 6\% for these cases.

For the case of $^{133}$Cs, the value of $\kappa_\text{hf}$ is about 40\% smaller than that
obtained in an earlier calculation \cite{BP:91a}
and slightly increases the size of the anapole moment of $^{133}$Cs
inferred from experiment \cite{wood:97,HAX:01}. Values of $\kappa_\text{hf}$
are also presented for the $7s-8s$ transition in Fr and
for microwave
transitions between ground-state hyperfine levels in atoms of potential experimental interest.

\begin{acknowledgments}
The work of W.R.J. was supported in part by National
Science Foundation Grant No.\ PHY-01-39928. U.I.S. acknowledges
support by Grant No.\ B516165 from Lawrence Livermore
National Laboratory. The authors owe thanks to V. V. Flambaum for calling
the early work of \citet{FK:85} to our attention.
One of the authors W.R.J. wishes to thank
B. R. Holstein for a useful discussion concerning constraints
on weak meson-nucleon coupling constants.
\end{acknowledgments}

\appendix
\section{Third-Order Perturbation Theory \label{A}}

We introduce a perturbation $V_I = H^{(1)} + H^\text{(hf)}$ into the many-body
Hamiltonian $H_0$ describing an atom and expand the many-body
wave function $\Psi$ of the bound state $v$ in powers of $V_I$
\[
\Psi = \Psi_v + \Psi^{(1)}_v + \Psi^{(2)}_v + \cdots
\]
to find
\begin{align}
\Psi^{(1)}_v &= \sum_{n\neq v} \frac{|n\rangle\langle n | V_I| v \rangle}
{E_v-E_n} \\
\Psi^{(2)}_v &= - \frac{1}{2}
\sum_{n \neq v} \frac{\langle v | V_I | n \rangle \langle n | V_I | v \rangle}
{(E_v-E_n)^2}\ \psi_v  \\
&- E^{(1)}_v \sum_{m\neq v}
\frac{|m \rangle\langle m | V_I | v \rangle}{(E_v-E_m)^2}
+
\sum_{\substack{m\neq v\\ n\neq v}}
\frac{|m \rangle\langle m | V_I | n\rangle
\langle n| V_I | v \rangle}{(E_v-E_m)(E_v-E_n)} ,
\end{align}
where
\begin{equation}
 E^{(1)}_v =\langle v | V_I| v \rangle \equiv \langle v | H^\text{(hf)} | v \rangle
\end{equation}
is the first-order correction to the energy.
The approximate wave function $\Psi_v + \Psi^{(1)}_v +  \Psi^{(2)}_v$
is normalized to second-order.

The third-order matrix element of the dipole
operator is given by
\begin{equation}
\langle w |z| v \rangle^{(3)} = \langle \Psi^{(2)}_w | z | \Psi_v\rangle +
\langle \Psi^{(1)}_w | z |\Psi^{(1)}_v \rangle +
\langle \Psi_w | z | \Psi^{(2)}_v\rangle .
\end{equation}
Expanding this expression, we obtain
\begin{multline}
\langle w |z| v \rangle^{(3)} = -\frac{1}{2}\sum_{m \neq w}
\frac{\langle w | V_I | m \rangle \langle m | V_I | w \rangle}
{(E_w-E_m)^2}\ \langle w | z | v \rangle
- E^{(1)}_w \sum_{m\neq w}
\frac{\langle w | V_I | m \rangle \langle m | z | v \rangle}{(E_w-E_m)^2}\\
+
\sum_{ \substack{m\neq w\\ n\neq w} }
\frac{\langle w | V_I | n\rangle
\langle n| V_I | m \rangle \langle m | z | v\rangle}{(E_w-E_m)(E_w-E_n)}
+ \sum_{\substack{m\neq w\\ n\neq v}}
\frac{\langle w | V_I| m \rangle \langle m | z | n \rangle
\langle n | V_I| v \rangle}
{(E_w-E_m)(E_v-E_n)} \\
-\frac{1}{2}\sum_{n \neq v}
\frac{\langle v | V_I | n \rangle \langle n | V_I | v \rangle}
{(E_v-E_n)^2}\ \langle w | z | v \rangle
- E^{(1)}_v \sum_{n\neq v}
\frac{\langle w | z | n \rangle\langle n | V_I | v \rangle }{(E_v-E_n)^2}\\
+
\sum_{ \substack{m\neq v\\ n\neq v} }
\frac{\langle w | z | n\rangle
\langle n| V_I | m \rangle \langle m | V_I | v\rangle}{(E_v-E_m)(E_v-E_n)}
\end{multline}
Setting $V_I = H^{(1)} + H^\text{(hf)}$ and retaining only those terms linear
in $H^{(1)}$, we obtain the expression given in Eq.~(\ref{eq2}).
It should be noted that the two terms above proportional to $\langle w |z| v \rangle$
do not contribute when the states $v$ and $w$ have the same parity.

\section{Angular Decomposition\label{B}}
The matrix element of the spin-independent operator $H^{(1)}$
between single-particle states $|i\rangle$ and $|j\rangle$ is
\begin{equation}
 \langle i | H^{(1)} | j \rangle = i\, \frac{G}{2\sqrt{2}} \ Q_W\
\delta_{\kappa_i\, -\kappa_j} \delta_{m_i\, m_j}
\int_0^\infty \!\! dr \, \left[ F_i(r) G_j(r) - G_i(r) F_j(r)\right] \, \rho(r) .
\label{a1}
\end{equation}
Here, $(\kappa_i,\, m_i)$ are angular momentum quantum numbers of the state $|i\rangle$
$[\kappa_i = \mp (j_i+1/2),\ \text{for}\ j_i=l_i\pm 1/2]$, $l_i$ and $j_i$ being the
orbital and total angular momentum, respectively, of the state $|i\rangle$. The functions
$G_i(r)$ and $F_i(r)$
are the large and small radial components, respectively, of the Dirac wave function
for the state $|i\rangle$.
We {\em define} the reduced matrix element of $H^{(1)}$ as the coefficient
of the angular momentum deltas in Eq.~(\ref{a1}). Using this (somewhat unconventional)
definition, it follows that
\begin{equation}
 \langle i \| H^{(1)} \| j \rangle = i\, \frac{G}{2\sqrt{2}}\ Q_W
\int_0^\infty \!\! dr \, \left[ F_i(r) G_j(r) - G_i(r) F_j(r) \right] \, \rho(r) .
\end{equation}

We decompose the spin-dependent operators of the type $H^{(k)}$,
with $k= (a,\, 2)$ in a spherical basis as\footnote{In this equation,
we omit the multiplicative factors
$\kappa_a$ and $\kappa_2$ defined in the Sec.~\ref{sec1} to avoid confusion
with the angular momentum
quantum numbers $\kappa_i$ introduced in the previous paragraph.}
\[
H^{(k)} = \sum_\mu (-1)^\mu\ I_{-\mu}\ K^{(k)}_\mu .
\]
The matrix element of the purely electronic operator $K^{(k)}_\mu$
between single-particle states $|i\rangle$ and $|j\rangle$ is
\begin{eqnarray}
\lefteqn{\langle i | K_\mu^{(k)} | j \rangle = i\, \frac{G}{\sqrt{2}}
\int_0^\infty \!\! dr \rho_v(r)\,
\Bigl[ \langle -\kappa_i m_i | \sigma_\mu | \kappa_j m_j \rangle F_i(r) G_j(r) \Bigr.}
\hspace{12em}\nonumber \\
&&\Bigl. -  \langle \kappa_i m_i | \sigma_\mu | - \kappa_j m_j \rangle\, G_i(r) F_j(r) \Bigr].
\end{eqnarray}
From this, it follows
\begin{eqnarray}
\lefteqn{\langle i \| K^{(k)} \| j \rangle = i\, \frac{G}{\sqrt{2}}\
\int_0^\infty \!\! dr \rho_v(r)\,
\Bigl[ \langle -\kappa_i \| \sigma \| \kappa_j \rangle F_i(r) G_j(r) \Bigr.} \hspace{12.5em}\nonumber \\
&&\Bigl. -  \langle \kappa_i \| \sigma \| - \kappa_j \rangle\, G_i(r) F_j(r) \Bigr].
\end{eqnarray}
Reduced matrix elements of the operator $\sigma$ are given by:
\begin{eqnarray}
\langle -\kappa_i \| \sigma \| \kappa_j \rangle &=&
(-1)^{j_i+\bar{l}_i-1/2} \sqrt{6\, [j_i][j_j]}\, \delta_{\bar{l}_i\, l_j}
\left\{
\begin{array}{ccc}
j_j & j_i & 1\\
1/2 & 1/2 & \bar{l}_i
\end{array}
\right\}
\\
\langle \kappa_i \| \sigma \| - \kappa_j  \rangle &=&
(-1)^{j_i+l_i-1/2} \sqrt{6\, [j_i][j_j]} \, \delta_{l_i\, \bar{l}_j}
\left\{
\begin{array}{ccc}
j_j & j_i & 1\\
1/2 & 1/2 & l_2
\end{array}
\right\} .
\end{eqnarray}
In the above, we have used the notation $\bar{l} = l(-\kappa)$ and $[j]=2j+1$.

The hyperfine operator is decomposed as
\[
H^{\text{(\text{hf})}} = \sum_\lambda (-1)^\lambda t_\lambda\ \mu_{-\lambda} \ ,
\]
where $\mu_\lambda = g_I \mu_N I_\lambda$ is the nuclear magnetic moment operator, and
$t_\lambda$ is the electronic part of the hyperfine interaction.
We may write the reduced matrix element of the magnetic moment operator
in the nuclear ground state as
\begin{equation}
 \langle I \| \mu \| I \rangle =  \sqrt{I(I+1)(2I+1)}\ g_I \ \mu_N \ .
\end{equation}
In the following,
the factor $\mu_N$ is
absorbed into the hyperfine interaction energy scale factor:
\[
W_{\text{hf}} = \frac{|e|}{4\pi\epsilon_0}\ \frac{|e|\hbar}{2M_p}\
\frac{1}{ca_0^2} = 1.987131\times 10^{-6} \ \text{a.u.}
\]
The electronic part of the
hyperfine interaction for a point nucleus in these units is given by
\begin{equation}
t_\lambda = -i\,\sqrt{2}\;
\frac{\bm{\alpha\cdot C}_{1\lambda}^{(0)} (\hat{r})}
{r^2} \ ,  \label{ax}
\end{equation}
where ${\bm C}^{(0)}_{1\lambda}(\hat{r})$ is a normalized vector spherical harmonic
\cite[p.\ 210]{var:88}.
For a distributed nuclear magnetization ${\bm M}(r)$, Eq.~(\ref{ax}) becomes
\begin{equation}
t_\lambda = -i\,\sqrt{2}\;
\frac{\bm{\alpha\cdot C}_{1\lambda}^{(0)} (\hat{r})}
{r^2} \ m(r) ,
\end{equation}
where the function $m(r)$ is given by
\[
m(r) = \frac{4\pi}{\mu} \int_0^r\!\! ds\, s^2\, M(s) =  \int_0^r\!\! ds\, s^2\, M(s)\ \div
 \int_0^\infty\!\! ds\, s^2\, M(s)\, .
\]
For the Fermi-type distribution given in Eq.~(\ref{rho}),
we find
\begin{multline}
 m(r,r<C) =  \frac{1}{\cal{N}} \left[ \frac{r^3}{C^3}
- 3\, \frac{ar^2}{C^3}\, S_1\left( \frac{C-r}{a} \right)
+6\, \frac{a^2 r}{C^3}\, S_2\left( \frac{C-r}{a} \right) \right.\\
 \left. -\ 6\, \frac{a^3}{C^3}\,  S_3\left( \frac{C-r}{a} \right)
+6\, \frac{a^3}{C^3}\, S_3\left( \frac{C}{a} \right) \right]
\end{multline}
and
\begin{multline}
m(r,r>C) =  1\ - \\
 \frac{1}{\cal{N}} \left[
  3\, \frac{ar^2}{C^3}\, S_1\left( \frac{r-C}{a} \right)
+6\, \frac{a^2r}{C^3}\, S_2\left( \frac{r-C}{a} \right) + 6 \frac{a^3}{C^3}\,
S_3\left( \frac{r-C}{a} \right)
\right]\, ,
\end{multline}
where  ${\cal{N}}$ is
\[
{\cal{N}} = \left[ 1 + \frac{a^2}{C^2} \pi^2 +
 6\ \frac{a^3}{C^3}\ S_3\left(\frac{C}{a}\right)  \right] \, .
\]
In the previous three equations,
\[
S_k(x) = \sum_1^\infty \frac{(-1)^{n-1}}{n^k} e^{-nx}\, .
\]

The reduced matrix element of the hyperfine operator $t$ is
\begin{equation}
\langle j \| t \| i \rangle = (\kappa_j+\kappa_i)
\langle -\kappa_j \| C_1 \| \kappa_i \rangle \
\int_0^\infty\!\! \frac{dr}{r^2}\ ( G_j(r) F_i(r) + F_j(r) G_i(r) )\ m(r) , \label{a2}
\end{equation}
where $C_{kq}(\hat{r})$ is a normalized spherical harmonic.
Finally, we note that the reduced matrix element of the dipole operator
$z$ is
\begin{equation}
\langle j\| z \| i\rangle = \langle \kappa_j \| C_1 \| \kappa_i \rangle\
\int_0^\infty \!\! dr \, r\, ( G_j(r) G_i(r) + F_j(r) F_i(r)) \ .
\end{equation}

With the aid of the above expressions, the reduced third-order matrix element
corresponding to Eq.~(\ref{eq2}) is found to be
\begin{eqnarray}
\lefteqn{ \langle wIF_w \| z \| vIF_v \rangle^{(\text{hf})} =
g_I \sqrt{I(I+1)(2I+1)[F_v][F_w]} \times } \hspace{0em}\nonumber\\[0.1ex]
&&
\Biggl\{ \sum_{j\neq v}
(-1)^{v-w+1}
\left\{
\begin{array}{ccc}
F_w & F_v & 1\\
j_j  & j_w & I
\end{array}
\right\}
\left\{
\begin{array}{ccc}
I & I & 1\\
j_j  & j_v & F_v
\end{array}
\right\}  \Biggr. \nonumber \\
&&\hspace{1em} \left[ \sum_{i}
\frac{\langle w\| H^{(1)}\| i\rangle\
\langle i \| z \| j \rangle\
 \langle j \| t \| v \rangle}
{(\epsilon_j-\epsilon_v)(\epsilon_i-\epsilon_w)}
+
\sum_{i}
\frac{\langle w\| z \| i\rangle\
\langle i \| H^{(1)} \| j \rangle\
 \langle j \| t \| v \rangle}
{(\epsilon_j-\epsilon_v)(\epsilon_i-\epsilon_v)} \right. \nonumber \\
&& \hspace{1em}\left. +
\sum_{i}
\frac{\langle w\| z \| j \rangle\
\langle j \| t \| i \rangle\
 \langle i \| H^{(1)}  \| v \rangle}
{(\epsilon_j-\epsilon_v)(\epsilon_i-\epsilon_v)}
- \frac{\langle w\| z \| j\rangle\
\langle j \| H^{(1)} \| v \rangle}{ (\epsilon_j-\epsilon_v)^2}
\langle v \| t \| v \rangle
\right]\nonumber \\
&&
\quad + \sum_{j\neq w}
(-1)^{F_v-F_w+1}
\left\{
\begin{array}{ccc}
F_w & F_v & 1\\
j_v & j_j  & I
\end{array}
\right\}
\left\{
\begin{array}{ccc}
I & I & 1\\
j_j & j_w & F_w
\end{array}
\right\}
\hspace{0em}\nonumber \\
&& \hspace{1em}   \left[ \sum_{i}
\frac{\langle w\| t \| j\rangle\
\langle j \| z \| i \rangle\
 \langle i \| H^{(1)} \| v \rangle}
{(\epsilon_i-\epsilon_v)(\epsilon_j-\epsilon_w)}
+
\sum_{i}
\frac{\langle w \| H^{(1)} \| i\rangle\
\langle i \| t \| j \rangle\
 \langle j \| z \| v \rangle}
{(\epsilon_j-\epsilon_w)(\epsilon_i-\epsilon_w)} \right. \nonumber \\
&& \hspace{1em}\Biggl. \left. +
\sum_{i}
\frac{\langle w\| t \| j \rangle\
\langle j \| H^{(1)} \| i \rangle\
 \langle i \| z  \| v \rangle}
{(\epsilon_j-\epsilon_w)(\epsilon_i-\epsilon_w)}
 - \langle w\| t \| w \rangle\
\frac{\langle w \| H^{(1)} \| j \rangle\
 \langle j \| z  \| v \rangle}
{(\epsilon_j-\epsilon_w)^2}
\right] \Biggr\} .
\label{eq2p3}
\end{eqnarray}

It is interesting to compare the interference term
with the the second-order reduced matrix element of
the dipole operator associated with the
spin-dependent terms $H^{(k)}$, $k=2,a$,
\begin{eqnarray}
\lefteqn{\langle wIF_w \| z \| vIF_v \rangle^{(2,a)} =
\sqrt{I(I+1)(2I+1)[F_v][F_w]}
  \times }\hspace{0.0in}\nonumber\\[0.2ex]
&& \sum_{j\neq v} \left[
(-1)^{v-w+1}
\left\{
\begin{array}{ccc}
F_w & F_v & 1\\
j_j & j_w & I
\end{array}
\right\}
\left\{
\begin{array}{ccc}
I & I & 1\\
j_j & j_v & F_v
\end{array}
\right\}
\frac{\langle w \| z \| j \rangle
\langle j \| K^{(k)} \| v  \rangle}
{\epsilon_v - \epsilon_j} \right.
\nonumber \\[0.5pc]
&&  \left. + (-1)^{F_v-F_w+1}
\left\{
\begin{array}{ccc}
F_w & F_v & 1\\
v & j & I
\end{array}
\right\}
\left\{
\begin{array}{ccc}
I & I & 1\\
j & w & F_w
\end{array}
\right\}
\frac{\langle w  \| K^{(k)} \| j \rangle
\langle  j \| z \| v \rangle}
{\epsilon_w - \epsilon_j}
\right] .  \label{eq3p3}
\end{eqnarray}
We find that the first term in this expression goes over
to the first term in the interference term under the replacement
\begin{eqnarray}
\lefteqn{\langle w \| z \| j \rangle \langle j \| K^{(k)} \| v  \rangle
\rightarrow} \hspace{0em} \nonumber\\
&& g_I \left[ \sum_{i}
\frac{\langle w\| H^{(1)}\| i\rangle\
\langle i \| z \| j \rangle\
 \langle j \| t \| v \rangle}
{(\epsilon_w-\epsilon_i)}
+
\sum_{i}
\frac{\langle w\| z \| i\rangle\
\langle i \| H^{(1)} \| j \rangle\
 \langle j \| t \| v \rangle}
{(\epsilon_v-\epsilon_i)} \right. \nonumber \\
&& \hspace{0em} \left. +
\sum_{i}
\frac{\langle w\| z \| j \rangle\
\langle j \| t \| i \rangle\
 \langle i \| H^{(1)}  \| v \rangle}
{(\epsilon_v-\epsilon_i)}
- \frac{\langle w\| z\| j\rangle\
\langle j \| H^{(1)} \| v \rangle}{(\epsilon_v-\epsilon_j)}
\ \langle v \| t \| v \rangle \right] ,
\end{eqnarray}
and that a similar correspondence can be made for the second term.
The completely different dependence on the intermediate state $j$
on the two sides of the above expression
explains the state-dependence of the coefficient $\kappa_\text{hf}$.

\end{document}